# Artificial Intelligence for the Internal Democracy of Political Parties

Claudio Novelli[1,2] · Giuliano Formisano[3,4] · Prathm Juneja[3] · Giulia Sandri[5] · Luciano Floridi[2,1]

## Abstract

The article argues that AI can enhance the measurement and implementation of democratic processes within political parties, known as Intra-Party Democracy (IPD). It identifies the limitations of traditional methods for measuring IPD, which often rely on formal parameters, self-reported data, and tools like surveys. Such limitations lead to partial data collection, rare updates, and significant resource demands. To address these issues, the article suggests that specific data management and Machine Learning techniques, such as natural language processing and sentiment analysis, can improve the measurement and practice of IPD.

## 1 Introduction

The robustness of democratic systems depends on the interplay of formal and informal elements. Formal mechanisms like constitutions, laws, and institutional design provide the foundational architecture. Informal elements, such as the media landscape, social norms, and the operation of political parties, play a crucial role in bolstering or undermining the democratic fabric. In this article, we focus on the operation of political parties and, more specifically, on their internal organisation

✉ Claudio Novelli
  claudio.novelli@unibo.it

1   Department of Legal Studies, University of Bologna, Via Zamboni, 27/29, 40126 Bologna, Italy
2   Digital Ethics Center, Yale University, 85 Trumbull Street, New Haven, CT 06511, USA
3   Oxford Internet Institute, University of Oxford, 1 St Giles', Oxford OX1 3JS, UK
4   Nuffield College, University of Oxford, New Road, Oxford OX1 1NF, UK
5   Centre d'Etude de La Vie Politique (CEVIPOL), Université Libre de Bruxelles, Avenue Jeanne 44, 1050 Brussels, Belgium

and what makes them democratic, often called Intra-Party Democracy (IPD). We examine existing methods for measuring IPD, question their efficacy, and explore the potential for enhancing them through Machine Learning (ML) techniques. We then turn our attention to ML's role in measuring IPD and helping parties in their day-to-day organisation.

Measuring IPD is challenging due to political groups' opacity, dynamism, and internal heterogeneity, which have long hindered research in this area. Recent advances in quantitative text analysis are providing new insights. Scholars analyse parliamentary speeches, scrutinise debates at party conferences, and pore over intra-party documents to assess ideological diversity. Additionally, they administer surveys and questionnaires to both party members and officials (Ceron, 2017; Bernauer & Bräuninger, 2009; Benoit & Herzog, 2017; Greene & Haber, 2017; 2017; Medzihorsky, Littvay, and Jenne 2014; Bäck, 2008). Digital technologies and social media websites offer fresh avenues for gathering relevant information to monitor and assess IPD. However, they also raise new questions about how they have reconfigured the dynamics of IPD itself (García Lupato and Meloni 2023; Dommett et al., 2021; Scarrow, 2013).

Existing methods for measuring IPD display some limitations. First, there is a degree of conceptual ambiguity surrounding what precisely constitutes democracy within a political party (Borz & Janda, 2020). Second, current metrics often focus on formal elements, such as party statutes, overlooking informal practices, like the influence of party factions or outside influences like trade unions. Third, standard empirical tools, like surveys and questionnaires, present multiple practical challenges, including limited data availability, social desirability bias, incapacity for regular updating, and high running costs.

In this article, we do not focus on conceptual ambiguity, an issue that affects all methodologies. Instead, we offer solutions to more practical challenges in IPD measurement given a specific framework of party organisation. To this end, we explore and map the applicability of data management and various ML techniques to IPD empirical measurement and research. These techniques span diverse tasks, from data collection and pre-processing to pattern recognition and quantitative measurement. We consider several ML techniques, e.g., automated text/data mining and natural language processing (NLP) (e.g., sentiment analysis, zero/few-shot classification[1]), classification algorithms (e.g., logistic regression), ensemble methods (e.g., random forest), and unsupervised learning (e.g., clustering algorithms).

Next, we analyse how political parties can leverage ML to improve the fairness or efficacy of their internal organisation and decision-making. Recent studies have shown that, especially in the EU, political parties increasingly use big data and digital technologies to campaign and run their organisational structures and functions (Barberà et al., 2021). Traditional European parties have progressively strengthened the use of digital technologies for internal functioning because of an external push factor (or contagion effect) in response to the rise of highly digitalised outside challengers, such as pirate or populist parties (e.g., Alternativet, Czech Pirate Party, Sumar, Five Stars Movement, etc.) (Jungherr et al., 2020). For instance, the new Synthetic Party in Denmark has used ML to elaborate its policy manifesto on the policies of Danish fringe parties since 1970–i.e., parties with a negligible share of the electorate–to reflect the interests and values of the 20% of Danish citizens who typically do not vote in elections. The Discord AI chatbot Leader Lars is its public face and figurehead.

---

[1] A zero-shot classification refers to practice of prompting a model to classify/label data without providing examples for several tasks, for example: tone, topic, questions-answers, and others. Alternatives include few-shot classification, when users provide models with some labelling examples, before employing the model to classify new/unseen data.

Parties are thus ready to integrate even more advanced techniques, such as ML, to improve their internal functioning. However, it is unclear how these techniques can be targeted to strengthen IPD and sustain parties' crucial linkage role with society. For this reason, in Sects. 3 and 4, we discuss ML techniques for more effective measurement of IPD, while in Sect. 5, we analyse the use of ML for enhancing IPD in practice.

The impact of ML on IPD is largely uncharted in academic studies. This article aims to fill this research gap by emphasising how data-driven tools could serve to monitor, assess, and improve IPD. For this analysis, we adopt a theoretical perspective and do not focus on any specific political group or context. However, while data management and ML offer powerful capabilities, some IPD challenges are inherently theoretical or political. For these issues, the technical solutions discussed here might be insufficient. Their effectiveness would likely depend heavily on the willingness of political actors (parties or the broader system) to address them.

The article is structured as follows. Section 2 introduces the concept of IPD, outlines its essential components and argues for the importance of democratic practices within political parties. Section 3 briefly analyses prevailing traditional methodologies to measure IPD and the obstacles they encounter. Section 4 explores the potential of ML techniques to address these challenges. Section 5 then pinpoints real-world applications of ML within political parties geared towards strengthening internal democracy. Section 6 concludes the article.

## 2 The Case for Intra-Party Democracy (IPD)

The configurations of intra-party organisations are many, multifaceted, and subject to frequent changes. Typically, parties evolve more rapidly than the regulatory context in which they operate.[2] A group of scholars offered an insightful framework for understanding and measuring intra-party organisation (Poguntke et al., 2016; Scarrow et al., 2017). They divide it into three primary dimensions: structure, resources, and representative strategies, further detailed in sub-dimensions, and used data gathered during the Political Party Database (PPDB) project to provide real-world insights into party life. Two contrasting metrics define each sub-dimension, representing opposite ends of a spectrum. For example, within the structure dimension, Centralisation and Localisation represent two extremes in terms of party structure and decision-making. The following framework, adapted from their work, will be used when discussing ML for practising IPD in Sect. 5.

---

[2] For this reason, some scholars use organizational theory to unpack this complexity, a discipline that shares analytical tools with party theory: e.g., strategic objectives, technological adaptations, and organizational culture (Borz and Janda 2020; Hatch 2018). A notable example of this approach can be found in the work of Kenneth Janda (Janda 1980; 1983).

(a) *Structure*. This dimension measures the party cohesiveness by pinpointing where and how decisions are made. It includes four subdimensions. Leadership Autonomy/Restriction (a1) describes how much a single leader can decide for the party and who can limit their power, such as a party board, members, or other elected figures. Centralisation/Localisation (a2) shows the balance of power within a party; highly centralised parties have top-down control, affecting candidate selection, party branding, and fund distribution. Coordination/Entropy (a3) considers how both internal (vertical) and external (horizontal) relations affect common action across the board. Territorial Concentration/Dispersion (a4) indicates a party's presence and organisation across a country's regions.

(b) *Resources*. This dimension concerns the distribution and use of financial and non-financial resources within political entities and their strategic significance. The subdimensions are broken down as follows. Financial Strength/Weakness (b1) compares a political group's economic resources against its rivals. Resource Diversification/Concentration (b2) identifies the variety in a party's funding sources. Parties dependent on a few major funders might prioritise those interests, whereas those with diverse small donations or volunteer support may focus on expanding their base engagement. State Autonomy/Dependence (b3) highlights a party's reliance on state funding, with high levels potentially pointing to a give-and-take relationship with voters, where tangible rewards are expected. Bureaucratic Strength/Weakness (b4) assesses the professional resources available to a party, suggesting that a robust organisational structure might influence party behaviour and reduce the need for volunteers. Volunteer Strength/Weakness (b5) looks at the human resources aiding in tasks like crowdfunding or campaigns.

(c) *Representative strategies*. This dimension concerns how parties determine and nurture relationships with their target audience. The subdimensions are delineated along the following lines. First, the individual linkage: Integrated Identity/Consumer Choice (c1) describes party efforts to bond with supporters; some parties prioritise memberships and shared activities to build a collective political identity beyond just championing policies. Others, especially personalist and populist parties, lean more towards presenting ideas without embedding a deep party affiliation. Second, the group linkage: Non-Party Group Ownership/Autonomy (c2) describes a party's ties to external entities. Indeed, some are primarily formed and driven by outside groups, like early trade-union-backed socialist parties, focusing on group rather than individual interests. Third, encapsulating the previous c1 and c2, the effect of the electoral formula (e.g., how votes are translated into seats) on IPD (c3) should be acknowledged. Following (Dow, 2010), in a majoritarian party system,[3] e.g., first-past-the-post, parties tend to cluster near the centre of the political spectrum. Thus, one would expect them to select moderate candidates. Instead, in proportional systems, parties support greater ideological dispersion. So, parties are more likely to elect candidates capable of differentiating themselves to win consensus.[4]

This empirically tested framework helps analyse how different parts of a political party change at varying rates. It also helps explore causal links between a party's internal organisation and its performance outcomes. For each subdimension, sample variables–e.g., party revenues for financial strength–inform potential measurement indexes (Poguntke et al., 2016; Scarrow et al., 2017).

Intra-Party Democracy (IPD) is fundamentally intertwined with the described framework, serving as both a driving force and an outcome of the various dimensions and subdimensions outlined. In this context, IPD refers to the distribution of power, resources, and decision-making procedures within a political party. It reflects a commitment to democratic norms and is both a driver of and an outcome of a more democratic party structure. While other definitions of IPD exist, this understanding aligns best with the framework and scope of our analysis.

However, an important question emerges: why should a political party be internally democratic? Parties seek to balance efficiency (timely decisions) with democratic practices (inclusivity, accountability). The US Democratic Party exemplifies this tension, prioritising primaries in 1972 but increasing leader influence later (Washington Post, 2021). Political parties are generally considered essential for democracy, but consensus is lacking on whether internal democracy within these parties is necessary. Some argue that external competition between parties is enough for democracy, and internal democracy might even weaken parties (Schattschneider, 1942; Dahl, 1970). IPD could also lead to susceptibility to outside influences or financial corruption (Close et al., 2019; Rahat, 2008). Despite these concerns, IPD strongly suggests that parties should reflect democratic values, particularly at the normative level (Dworkin, 1988). IPD promotes transparency and political accountability (Cross & Katz, 2013). Moreover, from a social perspective, IPD increases public trust through inclusive policymaking and candidate selection (Teorell 1999; Shomer et al., 2018). This fosters social cohesion and reduces political alienation. From an economic perspective, IPD may attract more donations due to transparency and fairer resource allocation. Many donors prefer giving to organisations that exhibit transparency and accountability, as they believe this increases their likelihood of having some influence over policy decisions, although this opens the potential risk of undue lobbying.[5] In functional terms, IPD strengthens parties eventually. Democratic parties are more adaptable and responsive to voters (Gauja, 2013; Rahat & Shapira, 2017).[6]

---

[3] By party-system, we employ Sartori's definition as the system of interactions resulting from inter-party competition (Sartori, 1976). Thus, we expect party systems to be comprised of two elements (i) the parties themselves (constituent units) and (ii) the interactions between them.

[4] Given the methodological nature of this paper, we do not explore further party-system considerations, which are beyond the scope of our investigation. However, within different electoral formulas, we point the readers towards other minor considerations: i) electoral tiers (i.e., at which levels votes are translated into seats), ii) ballot structures (i.e. how choices are presented to voters, e.g. voting for a candidate or party), iii) district magnitude (i.e. how many representatives are elected in a district), iv) legal thresholds.

While the arguments and incentives for IPD are compelling, some parties might resist adopting or strengthening it. They might have reasons to believe IPD is not in their best interest. However, this paper does not discuss the specific incentives that might persuade these parties to align with IPD standards. Instead, it primarily addresses parties that are inclined to proactively adopt some IPD practices, such as primaries. At the same time, it is reasonable to think that IPD assessments can be used as adversarial tools by competing parties, regardless of the intentions of the reluctant party. In fact, assessments may be entirely built on data not disclosed by the parties themselves. This means that even if parties do not want to implement IPD practices, there may still be a demand for measurement tools to evaluate them.

Given the value of IPD, how can one measure it, and what level of IPD could be considered satisfactory? We address these questions below, showing that current measurement methodologies have theoretical and empirical limitations.[7]

## 3 Measuring IPD: Current Methodologies and Their Challenges

Several scholars have proposed methodologies for measuring specific aspects of IPD (Bäck, 2008; Berge & Poguntke, 2017; Bille, 2001; Kenig, 2009; Rahat, 2009; Salgado, 2020). Other studies provide more comprehensive IPD indexes (Von Dem Berge et al., 2013; Rahat & Shapira, 2017; Scarrow et al., 2017). In this context, the Political Party Database Project (PPDB) fills a critical void by providing comprehensive cross-national data on political parties' formal structures and real-world practices (Poguntke et al., 2016; Scarrow et al., 2017).

The PPDB provides data on 410 political parties from 51 countries, from 2011 to 2014 in the first round and from 2016 to 2019 in the second round. Over 300 variables are documented, covering a broad spectrum of party functions from leadership selection and finances to manifesto construction and women's representation. These variables scrutinise the three primary dimensions of intra-party organisation, along with all the sub-dimensions previously outlined (Sect. 2): e.g., for Resources

---

[5] While some donors may prefer opaque political parties that conceal their financial support to candidates or parties, this approach is only viable in jurisdictions without mandatory disclosure requirements for funding and lobbying activities. And even in these cases, donors will likely demand some level of transparency regarding the party's internal operations, at least for themselves.

[6] This summary does not touch upon the supposed benefits of democratic organizations when it comes to the quality of their decision-making, such as those proposed under the Condorcet theorem. For a critique of this theorem, see (Austen-Smith and Banks 1996).

[7] Note that some of the challenges we discuss regarding the "computational" measurement of IPD can scale up to measuring party organisation in general. However, this broader analysis falls beyond the scope of our current investigation.

Diversification-Centralization, they analyse the ratio of public to private funding in a party's financial structure.[8]

Recently, Rahat and Shapira (2017) developed an IPD index employing empirical data and qualitative and quantitative measures. This index advances the measurement of IPD by analysing five dimensions–participation, representation, competition, responsiveness, and transparency–and uses a researcher-completed questionnaire, a cost-effective alternative to traditional surveys. Sources for the questionnaire include party documents, official communications, websites, and media coverage. Parties are rated on a 100-point scale across the dimensions, with the importance of each dimension dictating its weight in the overall score: 30% for participation, 20% each for competition and representation, and 15% each for responsiveness and transparency.[9] Based on their scores, parties are categorised as 'democratic' (61–100 points), 'partly democratic' (30–60 points), or 'non-democratic' (below 30 points).

Both the PPDB project and Rahat and Shapira's analyses greatly enrich the field of IPD research with their varied metrics and data.[10] However, they also face some difficulties.

### 3.1 Data Availability, Completeness, and Reliability

These datasets, while extensive, often rely on voluntary disclosures, self-reported data, and data from party members who choose to participate in surveys and interviews. This reliance presents challenges for data availability, completeness, and reliability. Availability may be compromised by parties' reluctance to share sensitive information, like staffing levels or minutes from internal meetings, either for privacy, competition or due to inconsistent record-keeping. The PPDB project, for instance, acknowledges the hesitancy of parties to report their number of payroll employees (Poguntke et al., 2016, 665), creating gaps that can alter organisational capacity assessments: selective disclosure or reporting and potential data loss further impact completeness. The variety in transparency and data handling across different countries contributes to inconsistencies in data quality. Reliability suffers from biases in survey and questionnaire responses provided by party members, with challenges like low participation rates, potential recall errors, social desirability biases, and observer effects. Furthermore, party members might intentionally provide misleading information to either score higher on IPD indexes or mislead rival parties about their strengths. These issues, compounded by selective disclosure, can skew the dataset, possibly misrepresenting the true extent of internal democracy within political parties. Similar constraints impact questionnaires completed by researchers or political analysts, as in (Rahat & Shapira, 2017), where the filling out of questionnaires can be significantly subjective.

---

[8] Additionally, this database supports the examination of various theories about party organization, leading to the development of distinct indexes, such as Assembly-based IPD (AIPD), Plebiscitary IPD (PIPD), and Open Plebiscitary IPD (OPIPD). AIPD evaluates how inclusive decision-making is regarding party policies, structure, and staff selection. PIPD measures the extent of one-member-one-vote practices for policy and personnel choices. OPIPD expands this to include non-party members (Berge and Poguntke 2017, 144). The project often employs logistic and multivariate statistical regressions to test hypotheses concerning party organization.

[9] The importance of dimensions has been assessed through a brainstorm section of the research group.

[10] We should clarify that the PPDB was initially designed to address party organization in general, rather than focusing specifically on internal party democracy (IPD). We thank one of the anonymous reviewers for correctly highlighting that some of our considerations might scale up to broader party organization, as IPD is a subset of this larger context. Nevertheless, it is important to note that the PPDB has previously been utilized for analysing and developing measurement methodologies pertaining to IPD (Berge and Poguntke 2017).

### 3.2 Updating and Monitoring

The frequency of updating the datasets is a critical limitation and is closely tied to the availability of resources. Although datasets may cover extended timeframes, updating these datasets is labour-intensive. This can lead to data not capturing swift transformations within party organisations, including reactions to electoral setbacks, leadership transitions, or policy shifts. Furthermore, for many variables, these datasets often capture only a single data point, which hampers the ability to trace the progression and internal dynamics of party structures over time.[11] For instance, if a party gradually shifts from a leader-centric model to a more member-driven approach over several years, the incremental nature of this transition may be obscured. Without real-time or annual updates, databases can miss short-lived but significant intra-party democratic experiments. An instance of this would be a political party exploring direct member policymaking through digital means for a short duration; such an initiative could remain unrecorded if it does not align with the data collection schedules.

These shortfalls exacerbate the challenges of achieving continuous monitoring. The database's difficulty in updating or reflecting a party's internal democratic evolution means it cannot provide real-time monitoring. Constant monitoring and more frequent data collection would allow for longitudinal studies, providing insights into how parties adapt to changing political landscapes, evolving social demands, and the impact of specific events and technological advancements. However, it is important to acknowledge that the parties themselves could consider much of this data confidential. Their willingness to share it will depend on their comfort level and relevant privacy regulations.

In this context, ML can be helpful for iterative analysis, but researchers will still need to perform crucial monitoring tasks on the ML-generated outputs, including plausibility checks.

### 3.3 Computational Effort

Maintaining a sizable database such as the PPDB incurs high costs and demands considerable time due to extensive data processing, analysis, and necessary updates. Management expenses encompass data collection, entry, quality assurance, and the computing infrastructure. The database's complexity demands advanced software and skilled analysts, constituting a significant investment that may limit update regularity and database expansion. This also holds for data obtained via questionnaires and analysed through coding systems developed after extensive brainstorming by research teams (Rahat & Shapira, 2017).

---

[11] For instance, in the PPDB (Poguntke et al., 2016, 662).

Developing a more profound, empirical analysis of IPD is particularly resource-heavy when it comes to statistical scrutiny. Computational costs can reduce the depth and regularity of analyses, risking a simplification of party democracy's evaluations. For example, limited computing power might prioritise quantifiable factors such as leadership candidate numbers over subtler elements like the inclusiveness of decision-making for rank-and-file members. Additionally, the vast amount of data combined with the necessity for precise documentation of multi-layered intra-party practices means verification and analysis can be slow. This lag can make the findings outdated, diminishing their relevance to current political debates. Thus, for instance, by the time an exhaustive study of gender balance in party leadership is completed, the parties in question might have already experienced further changes.

Computational costs significantly undermine the practical input of IPD indexes, especially for potential voters who require reliable information during election campaigns or voting. If the goal is to make IPD more than an academic exercise and integrate it into practical political engagement, the current limitations pose a serious obstacle to its real-world application and relevance.

### 3.4 Unmeasured or Opaque Variables

An additional point concerns the rigidity of the analysed variables. Many variables in the datasets relate to official documents, such as party statutes and regulations. While these documents are important for IPD as normative constraints that parties impose on themselves, there is a risk that real-world practices may significantly diverge from them. Take, for example, the much-debated superdelegate structure of the U.S. Democratic party in the 2016 and prior elections. In this case, 712 out of the 4,763 voting delegates who chose the party's nominee were 'unpledged' (i.e., untied to voter's preferences) (Stein, 2016). While this information is captured in party bylaws, the voting tendencies of those delegates are not; if the delegates tended to follow the voting patterns of the electorate, formal practices would understate the party's IPD, and if they tended to follow the wishes of party leaders, formal practices would overstate the party's IPD. Unofficial party subgroups, the cultural demands of a citizenry on a party to act democratically even if they don't necessarily have to, and the informal public power of party leaders are all examples of regularly non-formalised factors that may profoundly impact IPD. Several studies have demonstrated that party rules often differ from practices also in European parties. This is the case for open and democratic processes for selecting leaders and candidates: inclusive methods are often mandated by party bylaws but, in most cases, are either manipulated to fit the elite's needs or disregarded altogether (Cross & Katz, 2013). This unreliability should be considered alongside the empirical and practical limitations previously mentioned, as it can further complicate the accurate measurement of IPD.

As we shall argue in the next section, many of these difficulties can be removed or reduced by using data management and ML techniques to assess IPD, a strategy yet to be explored by relevant studies.

## 4 Machine Learning to Support IPD Measurements

Data management and ML techniques can be leveraged to address the four challenges previously identified and improve the measurement of IPD. In what follows, we link them to the internal organisational dimensions of political parties – structure, resources, and representative strategies, as detailed in Sect. 2.

### 4.1 Enhancing Data Availability, Completeness, and Reliability

ML can tackle the challenges of data availability, completeness, and reliability in measuring IPD. An essential technique for enhancing data availability is Natural Language Processing (NLP), which employs ML algorithms to interpret and extract meaningful information from vast amounts of unstructured text data which would otherwise be unusable. In the IPD context, NLP may extract insights from various text-based sources, such as public records, speeches, press releases, and social media (Grimmer & Stewart, 2013; Laver et al., 2003; Marwala, 2023). A prominent example of such NLP-based tasks could be to analyse non-textual data, which can be "prompted" into text. For instance, engagement data received by social media posts can be added to the posts themselves, so the machine uses them as extra contextual information. In other words, you would have a post's text, followed by information such as "this post received XX number of favourites, YY number of shares/retweets, ZZ number of replies/comments". This prompting exercise can increase the machine's performance as we provide contextual information. Similarly, weekly polling data can be added to press releases, etc. Models with image recognition capabilities may help convert previously machine-unreadable data into datasets suitable for model inputs. By doing so, NLP may infer pertinent information about party policies, leadership dynamics, and the degree of member participation. This approach compensates for the inherent scarcity of data about IPD and mitigates the impact of data withholding by political parties for reasons of confidentiality, although it may not fully compensate for all challenges posed by ambiguous language and context. Apart from prompting, ML outside of NLP methods may also be useful for combining disparate datasets by, for example, building a measure of member participation that includes attendance, social media, and traditional media measurements as variables. Moreover, ML enables the detection of hidden patterns and relationships in data that might escape human analysts, thus providing a deeper understanding of data's implications for IPD.

NLP also provides the framework and tools for sentiment analysis, potentially gauging public perception and internal sentiment regarding a political party's democratic nature, which may serve as a proxy for more direct measures of IPD (Mohammad, 2016). Sentiment analysis can be useful in extracting information from textual data (e.g., social media) that would otherwise be time-consuming for humans to annotate and then analyse (Ansari et al., 2020; Caetano et al., 2018; Hasan et al., 2018; Martínez-Cámara et al., 2014). Suppose, for instance, a political party has not disclosed detailed records of their primary elections, citing confidentiality. Sentiment analysis can be applied to social media discussions about the primary process among party members and followers. If the analysis reveals predominantly negative sentiments, especially regarding transparency and inclusiveness, it could suggest issues of IPD. Researchers could quantify these sentiments to create a sentiment score for each aspect of IPD, as well as study its variation over time and topics.

When limited to single data points, as often found in datasets like the PPDB, Predictive Analytics and imputation methods can extrapolate further data. This method uses historical data to estimate missing values where direct collection is unfeasible (Hastie, Friedman, and Tibshirani 2001). Suppose a party traditionally records the number of attendees at its annual meeting but fails to do so for the current year. However, the party has data on the number of attendees from previous years and knows that attendance spikes when there are hot-button issues on the agenda. If this year's meeting agenda included such issues, the party could use a regression or more advanced correlational model to estimate the likely attendance based on the correlation between agenda prominence and past attendance figures. The predicted attendance provides a (missing) data point that reflects member interest and engagement, which is a component of IPD. It is worth noting that imputation methods can be risky, as most models will be unable to account for all relevant variables for parties, such as members defecting to other political parties, changing political climates, and even the impact of weather on attendance at an annual meeting.

In short, training a correlational model on the historical attendance data makes it possible to understand the relationship between the variables (e.g., agenda prominence) and the attendance numbers. Such predictions, bolstered by techniques like ensemble methods and cross-validation (Dietterich, 2000), can serve as proxies for member engagement in IPD measurements.[12] In a similar vein yet distinct in application, the Data Imputation with the k-nearest neighbours (KNN) method addresses missing values by locating the 'k' closest data points and imputing values based on these (Batista & Monard, 2003). The 'k' neighbours must be carefully chosen to represent the broader dataset. For example, KNN would calculate the mean attendance from the most similar branches to estimate missing attendance at party meetings, determined by factors like location and size. This method preserves data uniformity internally without the need for external data sources.

Transfer Learning also offers a strategic advantage in contexts where data is limited. This technique involves repurposing a model created for a specific task to serve as the foundation for another (Pan & Yang, 2010). The performance of these models within political science has been analysed through comparative studies of various text classification techniques (Terechshenko et al., 2020). This approach is especially beneficial when data for the second task is scarce. In the context of measuring IPD, transfer learning might involve, for instance, fine-tuning a sentiment analysis model–initially trained to perform a generalist task (e.g., token or sentence prediction) on social media data from Country's party members–to evaluate sentiments in Country B, where the data is scarce (Kaya, Fidan, and Toroslu 2013).[13] Transfer Learning capitalises on the rich data insights from one region to bolster analysis in data-poor areas, thereby enriching the understanding of IPD across diverse landscapes.

---

[12] However, it is essential to acknowledge the risks of extrapolation and to understand the model's underlying assumptions when using regression models to predict missing data.

To improve data reliability in measuring IPD, we can use ML models for anomaly detection (Nassif et al., 2021; Omar, Ngadi, and H. Jebur 2013). These models can be designed to identify patterns that deviate from the norm, flagging outliers that may signify errors, manipulation, irregularities, or legitimate changes in behaviour that could represent positive developments in the IPD. It might work in the following way: a dataset of voting patterns across several internal party elections is analysed, including turnout, vote distribution, and spoiled ballots. The algorithm establishes a baseline for expected voting behaviour based on historical data. It then scans the current data for anomalies–such as an unexpected surge in turnout or unusual vote counts that starkly contrast with established trends. For example, if a party typically reports a 60% turnout and suddenly a 95%, anomaly detection could flag this as an outlier. Further investigation could reveal whether this was due to increased political engagement, an error in data reporting, unethical practices to inflate turnout figures, or legitimate innovations that deviate from historical trends. Such scrutiny could ensure that the data accurately reflects the party's democratic practices.

Finally, ML can bolster survey methodologies, enhancing data availability, completeness, and reliability for measuring IPD. ML can analyse historical survey data and identify the most predictive questions for measuring IPD in this context. This might improve the quality of data gathered and the reliability of survey-based assessments (Couper, 2013). Also, classification algorithms can predict which party members are less likely to participate in surveys based on past engagement data. To increase the response rates from these members, targeted communication strategies can then be developed. For instance, an ML model may help a political party refine its survey to gauge member views on electoral nominations better, as it can be prompted to act as a human member from a given area/region or demographic group. An ML model might find that questions about the clarity of the nomination process are strong indicators of the health of IPD. It may also predict low response rates among certain demographic factors (e.g., age, gender). This insight leads to tailored survey methods: online, mobile-friendly versions for young people and paper surveys for remote branches. Personalised reminders might be sent to those predicted to be non-respondents. This data-driven approach ensures a focused survey and broad participation, enhancing the quality and reliability of insights into the IPD.

---

[13] To ensure accuracy when transferring NLP models across different linguistic and cultural contexts, it's beneficial to incorporate local data to fine-tune the model.

However, with the recent closure of social media API for academics and researchers (e.g., Twitter/X in 2023) and real-time scraping or social listening being cost-intensive, data availability is becoming an issue. Researchers can explore alternative data sources to compensate for the loss of direct social media API access. These alternative sources include public forums, news aggregators or archives (Boumans & Trilling, 2016), specialised online communities like Reddit (Proferes et al., 2021), and other digital platforms like Google Trends (Prado-Román et al., 2021) that offer insights into public discourse and social trends.

### 4.2 Keeping IPD updated and monitored

Data management and ML techniques may enhance updating databases for IPD measurements by streamlining data collection and enabling robust time series and longitudinal analyses (Chatfield and Xing 2019; Nielsen, 2019). These methods help identify trends and patterns in IPD over time, monitoring the evolution of democratic indicators within parties and forecasting future developments. Specifically, techniques for data acquisition, such as automated data collection, web scraping, and real-time data streaming, can significantly enhance the process of updating IPD measurement databases. This method shines where data extraction requires discerning complex contexts or patterns, tasks which exceed the capabilities of basic rule-based systems (Warren & Marz, 2015). Both methods are crucial for compiling large datasets from which ML models can learn. It is possible to process information automatically from various platforms, like political party websites, social media, and press statements while addressing data quality and representativeness challenges. This minimises the need for labour-intensive methods, enabling datasets to be updated efficiently and accurately. Furthermore, ML systems may allow near-instantaneous dataset updates, adapting over time to new information (Box et al., 2015). These systems would continuously collect data and apply NLP to evaluate the textual data, employing tasks such as topic modelling, sentiment analysis, and named entity recognition. Consider, for example, the monitoring of intra-party elections. Tracking the occurrence of intra-party elections and member participation rates are all crucial for assessing IPD. ML systems can identify and harness data from digital platforms where intra-party elections use web scraping tools to gather information on election timetables, candidates, voter turnout, results, and member engagement. Subsequently, these techniques, including classification algorithms, are employed to analyse data–for example, to assess the competitiveness of electoral races by considering the number of candidates and margins of victory.[14]

In sum, by regularly collecting and analysing new data, such systems ensure IPD indicators are consistently updated with the most recent information on party activities (e.g., intra-party elections). The system can then leverage this processed data to detect and present trends over time–e.g., members' participation–using visual tools, providing stakeholders with timely updates and calling attention to trends, discrepancies, or noteworthy changes within a political party.

---

[14] An ML system can analyse not only quantitative data (e.g., turnout numbers) but also qualitative data (e.g., the sentiment of party members about the election process –Is this a qualitative data point? Not another predicted score by an AI? We should think of a more qualitative data point, for example: human coding to then train ML?).

Pattern Recognition and Classification may also support the updating and monitoring. ML identifies patterns within large datasets and studies connection and communication patterns between party members (e.g., network analysis) (Hastie, Friedman, and Tibshirani 2001). This is instrumental in tracing the evolution of party structures, pinpointing even the most nuanced changes that might elude human observers. Take, for example, the application of pattern recognition to scrutinise member engagement and voting behaviours. Traditional approaches, such as direct surveys, are labour-intensive and may fail to capture the dynamic nature of ongoing engagement. In contrast, ML-powered systems, or ensembles of them, may identify recurring engagement and voting patterns, offering a dynamic and comprehensive view of IPD. The process entails training an algorithm on a vast array of data points–from forum/meeting participation to policy debate contributions and party ballot votes–to identify indicators of engagement diversity. In other words, such systems can predict a score that reflects the party's performance on a specific aspect. Then, such a score is aggregated with other scores for other characteristics to ultimately produce a classification for each party on several dimensions. These indicators include potential spikes in activity levels preceding elections or important policy debates, consistent voting patterns on specific proposals, a broad spectrum of participation reflecting the party's demographics, and the overall sentiment in policy discussions. By recognising these patterns, the algorithm can notify analysts if a sudden drop in participation or a shift in the sentiment could indicate a problem with IPD. The algorithm is also continually retrained with incoming data, which helps prevent it from becoming less accurate over time due to model drift.

Finally, ML can enhance the updating and monitoring of IPD by employing predictive modelling (Box et al., 2015; Montgomery et al., 2012). This approach uses historical data and current trends to construct future scenarios for political parties. It is useful for anticipating how a party might react to significant events, like losing an election. Imagine a model that can predict a change in party leadership based on how members feel and how the party has performed in elections, especially if these factors match up with similar situations from the past. While predictive modelling yields provisional insights, it serves as an early alert system for possible shifts in IPD, enabling researchers and stakeholders to adapt proactively.

### 4.3 Decrease Computational Effort

Enhancing the process of data gathering and its accuracy, as well as improving the ability to refresh data and oversee the democratic features within political parties, must also be accomplished with greater computational efficiency than what is achieved with traditional methods. ML methods might play a pivotal role in mitigating the challenges associated with computational efforts in measuring IPD. A significant portion of the computational efficiency of computational techniques is attributed to their superior scalability, which ensures that a system can manage increasing workloads or expand to support growth without impeding performance (Bekkerman, Bilenko, and Langford 2011). Since political parties are subject to continuous change in their structure, membership, and procedures, databases must integrate new data types or regularly handle larger data volumes. As data grows in complexity and volume, the computational systems must scale in tandem.

Consider an IPD database, such as the PPDB, that begins with data on a handful of political parties' elections, candidate selections, and membership voting policies. Over time, as seen with the PPDB, it may encompass hundreds of parties, each with distinct practices and broader democratic measures like policy development, gender representation, and youth involvement. While traditional databases might struggle with the increased size and complexity, leading to processing delays, an ML-driven system can adapt through automated expansion, real-time learning, computational resource optimisation, and forward-looking analytics (Bertsekas, 2017). Also, automated data gathering and processing notably diminish the time and effort needed for these activities. With NLP and web scraping, an ML system can autonomously pull pertinent details from text, websites, and databases, circumventing manual data entry.

Additionally, ML algorithms might be trained to optimise the use of computational resources: they discern the most crucial data for IPD analysis, enabling smarter allocation of computational power and reducing superfluous processing. Techniques like Principal Component Analysis (PCA) streamline this process by distilling large datasets to their most significant features, simplifying the data's complexity for analysis. This dimensionality reduction helps remove irrelevant data while improving processes and enhancing model accuracy by preventing overfitting. While PCA is a long-used method in the social sciences, it and other dimensionality reduction techniques remain important and useful in more advanced ML pipelines. Consequently, models are generalisable and perform better on new, unseen data—advantages especially valuable in the extensive datasets encountered in IPD measurement. As Large Language Models (LLMs)[15] are becoming smaller, less computationally expensive and open-source models, such as Mistral-7B, are out-benchmarking larger models, signifying a trajectory where cheap-to-run models may fill many research needs (Jiang et al., 2023).

As an integral ML component, predictive modelling forecasts trends and patterns, moving beyond static data analysis (Hastie, Friedman, and Tibshirani 2001). Predictive models handle large volumes of data adeptly, potentially pinpointing key variables that affect IPD and projecting future developments within political parties. ML models with incremental or online learning can update their algorithms with new data without being entirely retrained, streamlining ongoing analysis. For example, Google's AI model, Gemini,[16] can retrieve live or old information, accessing a vast and up-to-date source of information through pre-loaded datasets and web documents. As a result, the need for repetitive data re-analysis diminishes as models can forecast based on existing data trends.

---

[15] These are powerful language-based algorithms trained on vast corpora of multimodal (texts, coding, images) datasets, such as Wikipedia, GitHub, and Google Scholar. These algorithms have often successfully passed college-entrance exams, medical and legal tests, as well as outperforming humans on several tasks (e.g., arithmetic, reading comprehension, knowledge-based classification, etc.).

[16] Google's AI, Gemini, available at: https://gemini.google.com/

Finally, ML further minimises errors that often accompany manual data processing. Automated analysis of survey responses ensures more precise and uniform outcomes. This automation advances accuracy and cuts costs by lowering the likelihood of error-driven revisions.

Regarding reproducibility and accessibility, our proposed pipeline meets the current scientific standards for findability, accessibility, interoperability, and reuse of digital assets. In particular, the models under analysis can be downloaded for free using the HuggingFace community,[17] where model specifications and updates are frequently reported and monitored. Moreover, when employing such models, researchers shall record their model version, seed to split data, as well as report the ratio (e.g., 70/15/15%, or 60/20/20%) split for training, evaluating and testing their classifiers, together with all the other set parameters (e.g. learning rate, early stopping, etc.). Finally, researchers can employ these models using free and online-available software, such as Google Colab, allowing storage and computational power to run several analyses via machine learning.

### 4.4 Measuring Opaque/Previously Ignored Variables

While researchers may still be limited in data availability, focusing on public statements, official documents, and information shared by parties, the ability of ML models to collect, measure, and analyse data at scale creates opportunities for researchers to examine previously understudied or ignored factors that impact IPD. Take the earlier example of party 'superdelegates' votes in selecting a party's candidate for an election. While very time-intensive, researchers could theoretically compare superdelegates' votes to the party electorate's votes and examine whether they map on correctly. It is even less likely, however, that researchers would be able to comb through the public statements made by each superdelegate to examine their reasons for voting for a specific candidate and check if those cited reasons were to increase IPD. Both tasks are trivial for an adequately trained ML model, improving researchers' ability to measure the actions and the stated intent of party members.

Public statements and social media posts are compelling data sources for ML models to detect trends in party activity. Data analysis, GraphML, and other network science ML modelling techniques can better examine the connections between party members or supporters, the flow of information (e.g., do party supporters tend to repeat the public statements of party leadership?), and even the number of factions within a party (e.g., network graphs of party leaders that follow and interact with other leaders and members). These measurements may give researchers better insight into the informal power structures within parties and how democratic those structures are in practice.[18] For example, Barberà et al. (2015) studied political communication among Twitter users, using network data to investigate the ideological positioning of voters (Barberá et al., 2015). Although not the scope of the paper, its findings show how within-party dynamics mutate over political topics. This information can, in turn, enable researchers to better understand party positioning in relation to their electorate. Similarly, Google Trends[19] is a valuable digital platform for gathering and monitoring data about politicians and parties. Google Trends can be used to study how much interest (i.e., a relative measure of search interest provided by Google for a given time and location) candidates received. Moreover, one may exploit trends' related keywords to study the keywords associated with a specific candidate, and subsequently, these can be used to study intra-party competition further. Work from Prado-Román et al. (2020) shows that Google Trends data provides useful information to predict election winners in the US and Canada. They find that there is a strong positive relationship between the interest received by candidates and their vote share in the months leading up to the elections.

---

[17] HuggingFace, accessible at: https://huggingface.co/

ML models, especially LLMs, may even analyse many news articles about parties and their members. These articles can help to understand more insightfully who has informal power within a party and if that maps to the formal power structure publicly presented by the party. The differences in how a party formally presents itself and operates in practice help researchers understand real-world IPD and detect discrepancies between public statements and practices, indicating transparency issues.

The advent of Transformer-based models (Vaswani et al., 2017) and API tools from OpenAI and similar providers have significantly enhanced NLP capabilities. For instance, GPT-4 can be fine-tuned for tasks like text classification and sentiment analysis or customised using Retrieval Augmented Generation (RAG) for more precise tuning. RAG enables the creation of specialised search summarisation engines tailored to a specific document set (Lewis et al., 2020). Integrating political party documents, surveys, and datasets into an LLM allows for efficient information retrieval with citations, streamlining various discussed applications.

In conclusion, ML is not meant to replace traditional methods of IPD measurement but rather to complement them. ML can enhance empirical analyses by providing data-driven insights. In the following section, we explore how to integrate ML techniques with the three-dimensional analysis of internal party organisation discussed earlier.

---

[18] GraphML Techniques: GraphML models are applied to analyse the network's structure. These techniques can uncover patterns of information flow (how information spreads through the network), the centrality of nodes (indicating influential figures within the party), and community detection (identifying subgroups or factions within the party).

[19] Google Trends, accessible at: https://trends.google.com/trends/

## 5 Real-world applications of Machine Learning for IPD

In Sect. 2, we outlined a framework for analysing and measuring political parties' internal organisation through three key dimensions and their respective subdimensions (Poguntke et al., 2016; Scarrow et al., 2017). Moving forward, we now illustrate how data management and ML can assist political parties in their daily organisational activities and enhance their democratic practices. Thus, while some points overlap with Sect. 4, this section focuses on measurement dimensions that can also provide effective recommendations for organisational change.

Recent research shows that parties are nearly always laggards when it comes to developing and using technology, especially outside the US. EU (and UK) parties lack the money and expertise to invest in data-driven techniques for campaigning or internal functioning (Dommett et al., 2024). This means that they often end up adopting overall inefficient systems and unsophisticated practices. Moreover, as mentioned above, using ML is risky for parties as it can make mistakes. Thus, it is mainly used for internal, administrative, and time-saving purposes, which means there is a huge potential for extending its use to a broader range of internal functions. We present the use of ML in IPD practices in alignment with the three dimensions of the PPDB main party organisation.

### 5.1 Structure

First, ML can be used to identify the topics a leader focuses on and how they evolve over time (Leadership Autonomy/Restriction (a1)). Understanding supporter sentiment and key issues can help tailor communication strategies to better resonate with the party base, leading to more effective engagement.

But how can this be achieved? Topic Modelling might gauge the emotional tone of a leader's communication, track the frequency and context of words related to power, decision-making, and autonomy, and compare leaders' public statements with official party documents to assess alignment. Predictive modelling can identify patterns in which party members' actions follow the leader's public statements. For example, consider how sentiment analysis can evaluate the tone and content of a leader's public addresses to determine how much freedom they have in their speech. If a leader's public statements significantly diverge from party policy or manifestos, this could indicate higher autonomy. Conversely, high alignment might suggest restrictions. Moreover, one could analyse a dataset of speeches from different party leaders over time. Using NLP, one could then identify changes in sentiment and topic adherence to party lines. The analysis could reveal if a leader expresses more personal opinions over time or becomes more restrained, indicating a shift in autonomy.

Second, political parties can leverage insights into the balance of power between central leadership and local branches to enhance their strategies and operational efficiency (Centralisation/Localisation (a2)). By understanding where the power lies within the party, leaders can make informed decisions about resource allocation, policy implementation, and member engagement.

What can ML perform in this regard? GraphML Techniques apply ML to graph-based representations of political party structures to discern and forecast decision-making dynamics. The process begins with collecting data on communications, financials, and decisions. Using graph algorithms, a network model of the party is constructed, pinpointing how power flows between nodes (individuals or branches). ML then scrutinises this network: it can show trends, like whether the party is becoming more centralised (power is getting more concentrated at the headquarters) or more localised (power is spreading out to regional offices or individual members). A more localised structure where many different people and offices have a say could indicate a more democratic setup. Alternatively, if just a few people at the top have all the power, it might be less democratic. Applying this model to proposed rules or structural changes may allow for better predictions of shifting power balances.

Third, Machine learning (ML) algorithms can calculate the entropy of decision-making data to measure the organisation and predictability of a party's decisions. Lower entropy values indicate concentrated and consistent decision-making, while higher values suggest randomness and unpredictability ((a3) Coordination/Entropy). Party leaders can gauge internal agreement or dissent by understanding their decision-making entropy. Also, ML models trained on historical voting records can predict proposed policy outcomes, aiding leaders in strategising and managing expectations.

For this purpose, Time-Series ML analysis can reveal the temporal patterns of decisions and actions, showing whether the party operates in a concentrated manner over time or displays spikes of entropy. Evidence can also come from the analysis of voting behaviours of party's members: the success rate of an ML model trained on historical voting records to predict outcomes based on established positions and past votes will reflect the concentration of decisions (predictable voting patterns align with the party line) versus entropy (varied and unpredictable voting). So, for instance, if the party's leadership proposes a new policy, and ML predicts voting outcomes based on historical alignment with such policies, a high accuracy would indicate a concentrated decision-making process. If the actual votes are highly variable and the ML predictions often fail, this shows a higher level of entropy, suggesting that individual members or factions within the party are making autonomous decisions, reflecting a more decentralised structure.

Fourth, ML algorithms can analyse geographical data to identify patterns in the distribution of a political party's influence and organisational presence (Territorial Concentration/Dispersion (a4)). Clustering algorithms can detect areas with high densities of party activities and membership and regions with sparse presence. By identifying areas with concentrated or dispersed activities, leaders can gauge their territorial influence and ensure they are engaging with a broad electorate, not just concentrated areas. Additionally, ML insights help allocate resources and plan events more effectively, targeting areas needing more attention.

For this purpose, ML algorithms can process geographical data to identify patterns in the distribution of party branches, events, and membership; clustering algorithms can detect areas with high densities of party activities versus those with sparse party presence. So, to determine whether a party's influence is centralised in some areas or is effectively reaching out to diverse regions, an ML algorithm can analyse location data and understand whether some regions have higher concentrations of resources and activities. By contrast, if ML finds that party activities and resources are spread across different regions, this suggests territorial dispersion, implying that the party is trying to be inclusive and democratically engage with a broader electorate. Training these models on existing data will make it possible to prioritise planning future events in historically neglected areas, potentially enhancing IPD.

### 5.2 Resources

As already seen, how a party manages its resources can affect its operations, strategies, and, ultimately, its democratic nature. ML analysis of financial resources plays a significant role in assessing IPD, as economic resources influence a party's ability to campaign, set agendas, and implement policies. When direct data is unavailable, ML algorithms can rely on proxy indicators such as publicly accessible election spending records. Comparative analysis using financial data from similar parties can also provide insights, allowing for informed estimates of a party's financial circumstances.

In this context, ML can first analyse comprehensive financial data, such as annual reports, donation records, campaign expenditures, and debts or loans, to evaluate a political party's financial strength and resource diversification ((b1) and (b2)). Parties may benefit from this in many ways: e.g., to identify risks from over-reliance on a few large donors and predict financial stability; to forecast cash flow trends and detect unusual financial transactions, helping prevent mismanagement.

For this purpose, clustering algorithms categorise donors based on their donation size and frequency, thereby revealing any concentration of funding within specific groups. This analysis can help analysts – and parties themselves–assess their democratic standing and identify areas where parties should diversify their fundraising strategies to mitigate potential risks. However, ML can also *predict* financial stability: it can predict cash flow trends and detect unusual financial transactions or changes in spending patterns that could indicate financial mismanagement or imbalances in resource allocation, recognising patterns in fundraising activities, donor contributions, and expenditure trends, ML may provide a comprehensive view of the party's financial operations. So, for instance, imagine a political party with multiple sources of income, including donations, government funding, and membership fees. In this context, predictions of stable finances suggest the party has the strength to support democratic activities like campaigns and policy development. Also, consistent donation patterns from diverse sources support financial independence, which is conducive to IPD.

Second, ML may be useful in assessing political parties' state autonomy or dependence by analysing various factors such as membership size, ideological stance, and electoral performance (State Autonomy/Dependence (b3)). Specifically, ML algorithms can uncover patterns and relationships that might not be immediately apparent through traditional analysis methods. By understanding their reliance on state funding, parties can strategise to mitigate or leverage this dependence more effectively.

For instance, regression analysis (e.g., non-/linear models, XGBoost and Random Forests) can explore the relationship between a party's policies and reliance on state funding. If a party's policy changes correspond with fluctuations in state funding, revealed through ML analysis, this could indicate a concerning level of dependence on state support. Also, the ML model might reveal that parties with a strong base of paying members are less reliant on state funding, whereas parties that struggle to attract members depend more on state support. Similarly, the model could show that parties with more extreme ideologies might find it harder to raise funds from private donors, making them more dependent on state resources.

### 5.3 Representative Strategies

Political parties may need to understand how effectively they foster a sense of collective identity versus catering to individual preferences (Integrated Identity vs Consumer Choice ($c_1$)). Identifying member engagement and retention trends is crucial for maintaining a strong, active membership base. By understanding the balance between collective and individual language, parties can tailor their communications to resonate more effectively with members. By understanding the balance between collective and individual language, parties can tailor communications to better resonate with members. Identifying engagement and retention trends helps parties strengthen member loyalty and participation.

Data analysis and ML algorithms can analyse party communications and member interactions in this context. For example, using NLP, one can quantify the frequency and context of collective identity markers (like "we", "us", "our party") versus individual consumer choice markers (like "you", "your choice", "your policy"). Also, ML analyses membership data to identify engagement and retention trends, which indicate the strength of integrated identity. By evaluating the sentiments expressed by party members and supporters on social media, ML can infer the emotional connection that individuals have with the party and party supporters' demographic data from their social media behaviour (e.g., m3inference by Wang et al., 2019). So, for instance, by scraping social media to analyse the language used by party members, an ML model could classify posts as reflecting either a collective identity or individual consumer choice. Posts that discuss shared values and group activities might be tagged as 'collective identity'. In contrast, those that focus on policy preferences without reference to group identity could be tagged as 'individual consumer choice'.

Finally, one of the primary challenges in IPD analysis is understanding the nature and strength of connections between political parties and external organisations, such as unions or NGOs. The relationships can range from non-party group ownership, where external organisations heavily influence the party, to complete autonomy, where the party operates independently (Non-Party Group Ownership to Autonomy ($c_2$)). Understanding the influence of external organisations is crucial for both political parties and external observers. This insight

**IPD via ML and data management - Summary**

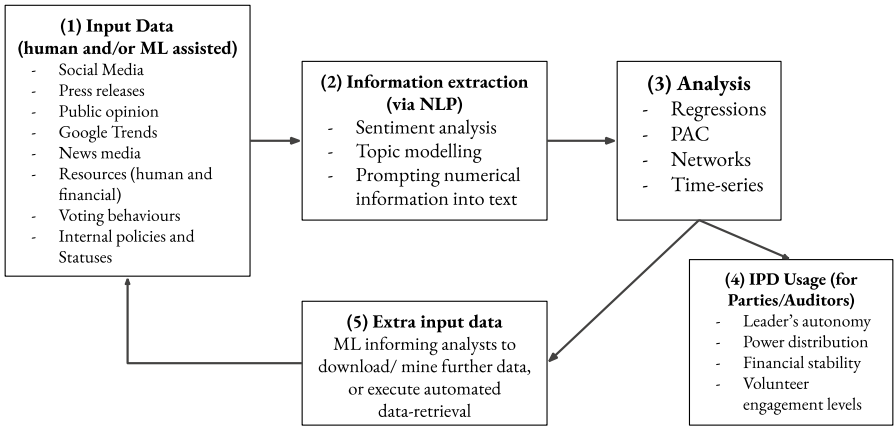

**Fig. 1** IPD via ML and data management—Summary (This plot is inspired by (Jin and Mihalcea 2023, 147).)

helps parties strategise alliances, prioritise policies, and craft communication strategies. By continuously monitoring their relationships with external groups, parties can ensure they maintain desired levels of autonomy or manage the extent of external influence effectively. For external observers, highlighting these relationships holds parties accountable for their alliances and policy origins. Consequently, voters can make more informed decisions by knowing which external organisations influence the parties they support.

For this purpose, ML can map the networks and interactions between parties and external organisations, such as unions or NGOs, and the strength and directionality of these links. Moreover, ML can examine the timelines of policy changes and external group activities to determine if there is a causal relationship, suggesting Ownership more than Autonomy. For instance, an ML-powered network analysis can be used to study the affiliations between a party and trade unions. By examining the co-occurrence of party policy announcements and union activities, one could assess whether the union's actions precede and possibly influence party policies, indicating a 'non-party group ownership' scenario. Conversely, a more autonomous party might show policy changes that are not closely followed by or aligned with any external group's activities.

Figure 1 summarises our workflow for IPD via ML and data management. It shows our procedure from (1) input data, passing through (2) information extraction, to (3) analysis, and then to either (4) ending the workflow (IPD usage), or going back to (5) data retrieval and re-starting the cycle. These procedures largely mirror IPD assessment without ML, but the content and data used differ.

# 6 Conclusion: risks of using Machine Learning for IPD

In this paper, we argued that ML and data management techniques can improve how we measure and then practice IPD. Current methodologies for measuring IPD rely on limited data sources, infrequent updates, and subjective interpretations, often leading to incomplete and unreliable data.

ML offers a solution by leveraging various techniques such as Natural Language Processing (NLP) and sentiment analysis. NLP can extract data from unstructured sources like speeches, social media posts, and news articles. We illustrated that sentiment analysis can then be used to gauge public perception and internal party dynamics regarding democratic practices. Additionally, predictive analytics can address missing data points by using historical information and established correlations.

By incorporating ML, researchers and political parties themselves can gain a more comprehensive and up-to-date understanding of IPD. This can lead to several benefits: improved IPD measurement, enhanced transparency (as ML can help reduce reliance on self-reported data from parties), real-time monitoring, and data-driven decision-making.

However, it is important to acknowledge ML's limitations in this context. Data availability and party willingness to share information are still significant hurdles. NLP and other ML techniques are also susceptible to biases and other inaccuracies (e.g., hallucinations in large language models) that must be addressed through specific technical and normative standards (Ziosi et al. 2023). While challenges remain, ML offers a promising avenue for more effective measurement and improvement of IPD within political parties.

**Acknowledgements** CN's contributions were supported by funding provided by Intesa Sanpaolo to the University of Bologna.

**Funding** Open access funding provided by Alma Mater Studiorum - Università di Bologna within the CRUI-CARE Agreement.Funding  Open access funding provided by Alma Mater Studiorum - Università di Bologna within the CRUI-CARE Agreement.

**Publisher's Note** Springer Nature remains neutral with regard to jurisdictional claims in published maps and institutional affiliations.